# Improved Algorithm for Throughput Maximization in MC-CDMA

Hema Kale[1], C.G. Dethe[2] and M.M. Mushrif[3]

[1] ETC Department, Jhulelal Institute of Technology Nagpur, India.
Email: `hema.kale72@gmail.com`
[2] ECE Department, Priyadarshni Institute of Engineering and Technology, Nagpur, India.
Email: *cgdethe@yahoo.com*
[3] ETC Department, Yashwantrao Chavan College of Engineering, Nagpur, India
Email: `milindmushrif@yahoo.com`

## ABSTRACT

*The Multi-Carrier Code Division Multiple Access (MC-CDMA) is becoming a very significant downlink multiple access technique for high-rate data transmission in the fourth generation wireless communication systems. By means of efficient resource allocation higher data rate i.e. throughput can be achieved. This paper evaluates the performance of group (subchannel) allocation criteria employed in downlink transmission, which results in throughput maximization. Proposed algorithm gives the modified technique of sub channel allocation in the downlink transmission of MC-CDMA systems. Simulation are carried out for all the three combining schemes, results shows that for the given power and BER proposed algorithm comparatively gives far better results.*

## 1. INTRODUCTION

Resource allocation is a major issue in the performance of wireless networks. Out of the available resources frequency allocation is the problem which is going to be considered in this paper. Channel fading is different at different sub carriers, this feature can be exploited for allocating the subcarriers to the users according to the instantaneous channel state information (CSI).The SCS-MC-CDMA system as shown in Fig.1 assigns to each user a selected number of sub-carriers [14]. The concept of sub-carrier selection is introduced to counter the problem of high power consumption. MC-CDMA systems usually have lots of sub-carriers, by using the appropriate sub-carrier selection technique, it can be made possible to match the power consumption to the user's data rate.

Sub carriers are allocated depending on user's data or on instantaneous Channel State Information (CSI). Instantaneous CSI refers to the amount of channel fading user experiences on particular channel. Some schemes have been proposed in already published papers for subcarrier selection according to CSI which includes,

- Selecting the subcarrier receiving maximum power on it.
- Selecting the sub-carrier with maximum SNR.

- Selecting the subcarrier requiring least amount of transmit power on it.

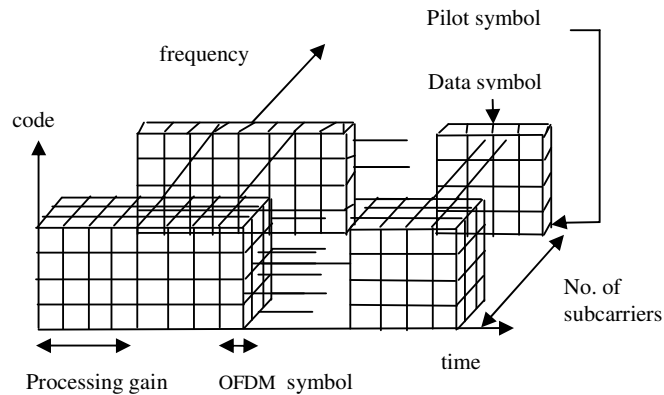

Figure.1 Frame format of sub carrier selecting MC-CDMA system .

Out of which last one method has been implemented for achieving optimum throughput in this paper.

An appropriate sub-carrier selection technique results in high spectrum efficiency, reduction in high power consumption at the mobile terminal, high data throughput in a multicell environment, improvement in BER performance, reduction in signal processing at the mobile terminal.

For the given power, throughput can be maximized by assigning maximum number of subcarriers to the users.

Here some throughput maximization techniques are given. The maximum throughput of each code channel can be achieved by employing following techniques,

- If the code channel is allocated to one user.
- If the code channel is allocated to the user who has the highest achieved SINR value with the same transmit power.
- If all code channels of one group are allocated to only one user.
- If the allocated transmit power of the one particular group is equally distributed over all code channels of such group.
- If the transmit power over different groups is allocated by water filling method.

In [13] Qingxin Chen, Elvino S. Sousa and Subbarayan pasupathy proposed a Water-filling algorithm , it was motivated by the water-filling (WF) principle in information theory i.e. given parallel channels with Gaussian noise, information should be first fed into channels with lower noise levels to achieve the maximum channel capacity. it improves the speed and average SINR of the system. One pitfall might lie in the case where one user's fading amplitudes are much larger than the average. G.K.D.Prasanna Venkatesan, and C.Ravichandran, has suggested a dynamic sub-carrier allocation technique for adaptive modulation based MC-CDMA system in [5] which results in improvement in throughput and BER performance. In this paper water filling algorithm is used to select the best sub-carrier,over the existing subcarriers. In this the principle of adaptive modulation consists of allocating many bits to carriers with a high SNR, whereas on carriers with low SNR only a few or no bits at all are transmitted. However there will be a possibility that when many channels suffer deep fading then there will be no transmission or very few bits are transmitted. In [15] the random policy has been described. This random policy is to allocate the subcarrier randomly regardless of channel information, which is concluded in [15] to be the best algorithm to get better BER performance. However the increase in system throughput with increase in SNR is less as compared to other advanced systems of sub carrier allocation. In [7] Tallal El Shabrawy1 and Tho Le-Ngoc2, has proposed a Subcarrier group assignment for MC-CDMA wireless etworks resulting in high throughput. In this paper, two interference-based subcarrier group assignment strategies are introduced for multicell MC-CDMA systems. In least interfered group assignment (LIGA), users are assigned to groups experiencing minimum interference. In best channel ratio group assignment (BCRGA), the user is assigned to the subcarrier group that holds the best ratio of channel response-to-received interference. In [14] Teruya Fujii, Noboru Izuka, Hiroyoshi Masui, and Atsushi Nagate, has proposed a sub-carrier selecting MC-CDMA system for 4G systems in which the concept of sub-carrier selection to counter the problem of high power consumption has been introduced. In this paper the concept is to assign each user only as many sub-carriers as are needed to support the user's data rate. In this method the system complexity increases when more number of filters are required for subcarrier selection.

In this paper we investigate the method of subchannel allocation to the user for the given transmit power in the downlink transmission. First the available channels are divided into number of groups and then this groups are allocated to the users. According to CSI each user will require a different transmit power on each channel, using this characteristic group of channels will be allotted to users. In the proposed algorithm the method of group allocation to the users has been modified which will result in producing global minima. This will result in further saving of the power and achieving higher throughput as compared with the original algorithm.

The rest of the paper commences with section II giving simplified introduction of original algorithm. In section III an improved method of subchannel allocation to the users is proposed. Then simulation results are discussed by considering one specific environment of CDMA network in section IV. At the end conclusions are presented.

## 2. Original Algorithm: Adaptive Sub-Carrier Allocation

This algorithm as discussed by Jun-Bo Wang, Ming Chen and Jiangzhou Wang [1] is focuses on the joint channel and power allocation in the downlink transmission of multi-user MC-CDMA systems for throughput maximization, under the constraints that the total transmit power should not exceed the maximum transmit power and each channel's SINR should not be less than a pre-defined value.

If the required amount of transmit power of each channel has been determined for all users before the channel allocation, then throughput maximization problem is given by a following optimization of $c_g^u$ problem as,

$$\max \sum_{u=1}^{U} \sum_{g=1}^{G} c_g^u \qquad (1)$$

Where
$c_g^u$ - number of the $u^{th}$ user's channels on the $g^{th}$ group.
**U** - Total number of users
**G** - Total number of groups of subcarriers.

Problem (1) is subject to

$$\sum_{u=1}^{U} \text{sgn}(c_g^u) \leq 1, \quad \forall u, g \qquad (1.a)$$

$$\sum_{u=1}^{U} \sum_{g=1}^{G} c_g^u p_g^u \leq P_T^{max} \qquad (1.b)$$

$$c_g^u \in \{0, 1, \ldots, S\}, \quad \forall u, g \qquad (1.C)$$

Where
**S** – Total number of subcarriers in $g^{th}$ group.
Above equation (1.b) is the total transmit power constraint.
where
**$P_T^{max}$** - The maximum transmit power, and
**$p_g^u$**- The required transmit power for $u^{th}$ user on one channel of the $g^{th}$ group, it is expressed as,

$$p_g^u = \beta N_o S^{-2} \sum_{s=1}^{S} |\omega_{g,s}^u|^2 \sum_{s=1}^{S} |\omega_{g,s}^u f_{g,s}^u|^{-2} \qquad (2)$$

Where

**$\beta$-** Target threshold of BER.
$f_{g,s}^u$ – $u^{th}$ user's channel fading on the $s^{th}$ subcarrier of the desired group
$\omega_{g,s}^u$ - $u^{th}$ user's frequency domain combining weight for the signal on the $s^{th}$ subcarrier of the desired group
Denoting $u_{gmin}$ as the user whose transmit power for one channel on the $g^{th}$ group is minimum among all users, i.e.

$$u_{gmin} = \arg\min_{1 \leq u \leq U} \{p_g^u\} \qquad (3)$$

## 2.1 Criteria used for group allocation

In the original scheme [1] the criteria used for group of channels allocation can be defined as, each group is assigned to the user who requires the minimum transmit power for one channel on that group. As shown in fig. 2 group allocation starts from first group.

In the original scheme the subcarriers are allocated to the different users by,

1) Calculating required power on each channel of all groups.
2) A group is assigned to the user who requires minimum transmit power for one channel on that group,
3) Starting from the first group, it will scan all the U no. of users and the user requiring minimum power will be allocated the first group.
4) Next second group will scan remaining (U-1) users, and the user requiring min. power will be allocated the second group and so on………

Refer flowchart for the above adaptive channel allocation algorithm which is constructed as below in fig. 2, it was not given in the earlier paper,

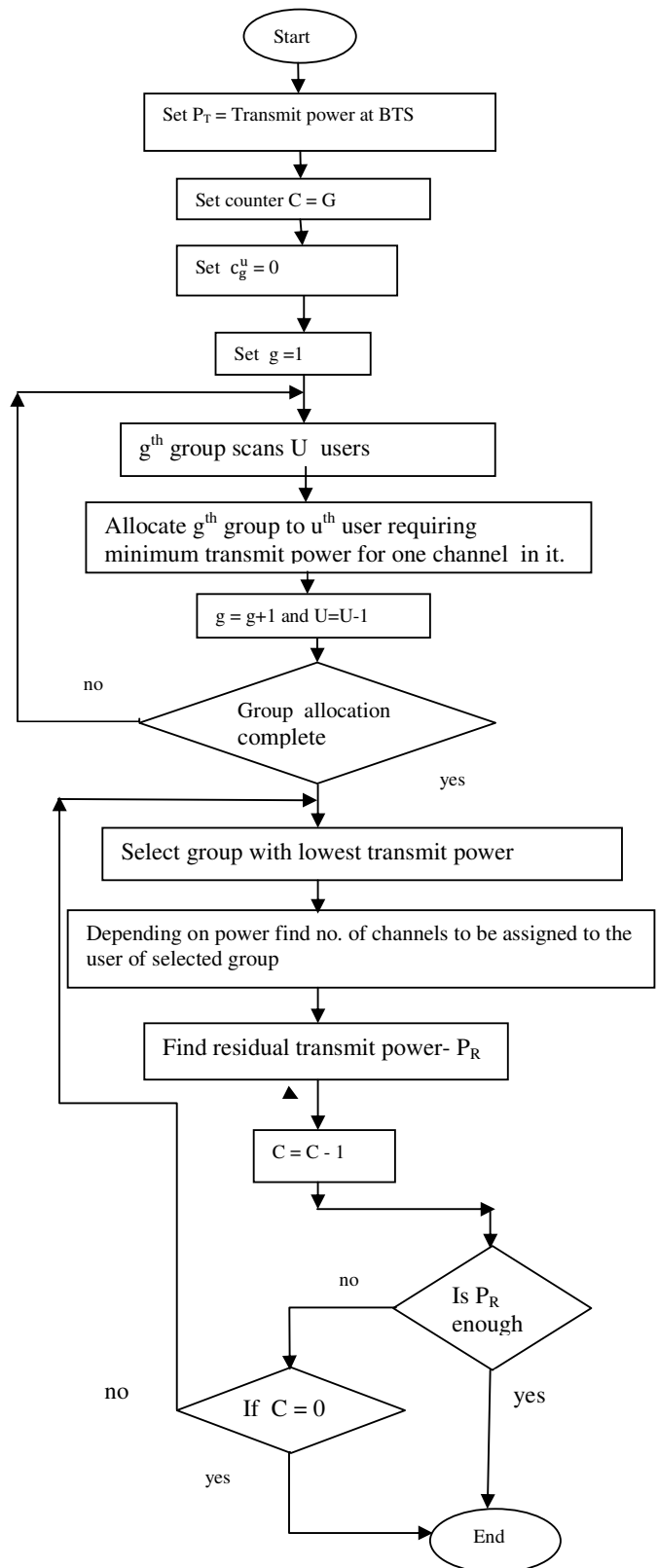

Figure. 2 Flowchart for original algorithm

## 3. Proposed Improved Algorithm For Subchannel Allocation

An improved algorithm is proposed for the channel allocation in the downlink transmission of multi-user MC-CDMA systems for throughput maximization, under the constraints that the total transmit power should not exceed the maximum transmit power and each channel's SINR should not be less than a pre-defined value.

In the Proposed algorithm group assignment technique is modified by finding out the global minima while allocating groups to the users. The subsequent chnnel allocation within the selected group is the same as in the original algorithm [1].
The proposed algorithm is explained as given below,

### 3.1. Criteria used for group allocation

In the proposed modified scheme, the group of subcarriers are allocated to the different users by,

1) Calculating required transmit power on each channel of all groups.

2) while allocating groups to the users, all the M number of groups will scan all the K number of users at the same time and the user requiring minimum transmit power on any one channel of any group will be allocated that group.

3) Next remaining (G -1) number of groups will scan all the remaining (U -1) number of users and so on…….

The proposed improved algorithm is as follows

**Initialization**

$P_R = P_T^{max}, C = \{1,2,…,G\}, c_g^u = 0$ for

$u = 1,…,U$ and $g = 1,…,G$.

**Group assignment**

**while** $C \neq \emptyset$

$u = 1:U$

$g = 1:G$

$[p_{min}, u_{gmin}] = \min(\min\{p_g^u\})$   % allocate a group to user
        requiring least power on one channel of that group

**end**

**Channel allocation**

**while** $C \neq \emptyset$

$t = \arg\min_{\forall g \in C} \{p_g^{u_{gmin}}\}$; % select the group with lowest power requirement

$c_t^{u_{tmin}} = \min\left(\left\lfloor \frac{P_R}{p_t^{u_{tmin}}} \right\rfloor, S\right)$; % calculate the available channel number

$P_R = P_R - c_t^{u_{tmin}} p_t^{u_{tmin}}$ ; % calculate the residual transmit power

$C = C \setminus \{u_{tmin}\}$ ;

**If** $c_t^{u_{tmin}} = 0$ % since the residual transmit power is not enough, terminate channel allocation.

Break the loop;

**End If**

**End While**

Denoting $u_{gmin}$ as the user whose transmit power for one channel on the **g**<sup>th</sup> group is minimum among all users.

A flowchart is constructed for the proposed improved algorithm as given below,

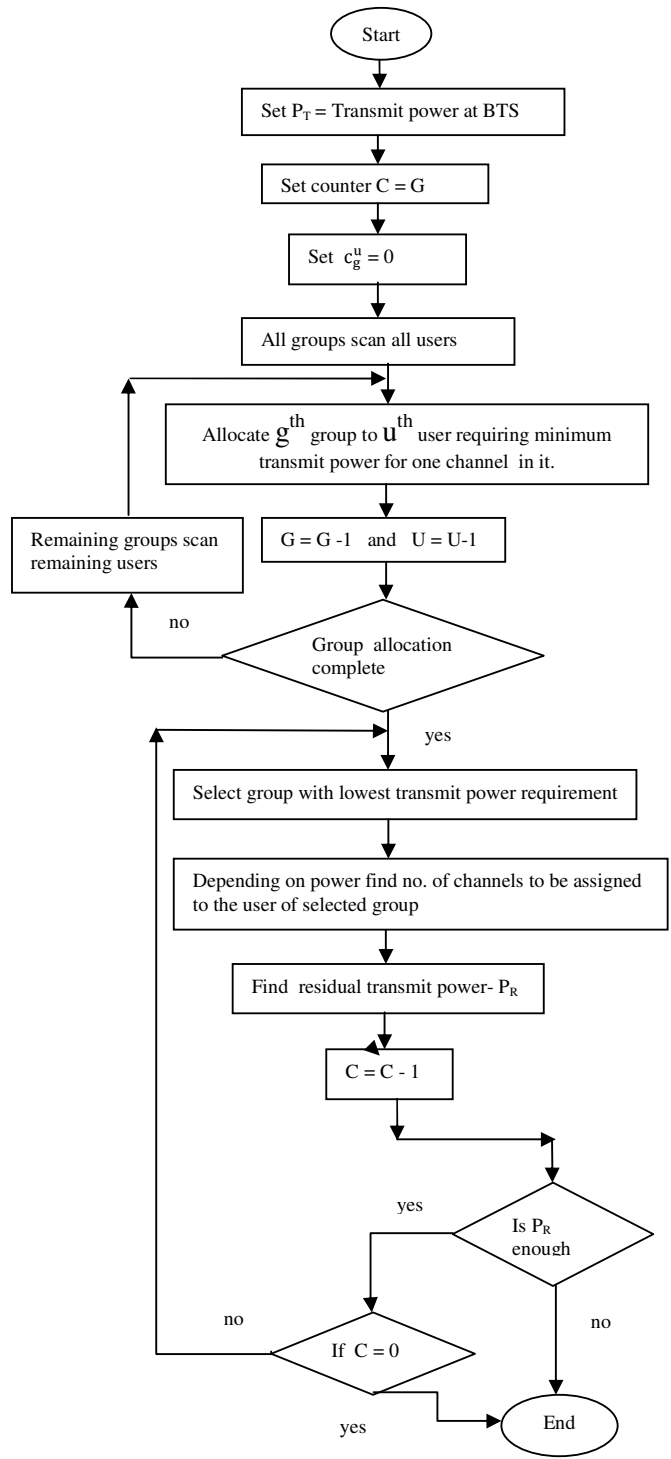

Figure. 3 flowchart for proposed algorithm

Different combining schemes will result in different power allocation, accordingly required transmit power ($p_m^k$) will change as per Table1[1] . Therefore throughput will be different for different combining schemes.

TABLE I..
REQUIRED POWER FOR COMBINING SCHEMES

|  | MRC | EGC | ZFC |
|---|---|---|---|
| $p_g^u$ | $\beta N_o S^{-2} \sum_{s=1}^{S} \lvert f_{g,s}^u \rvert^2 \sum_{s=1}^{S} \lvert f_{g,s}^u \rvert^{-4}$ | $\beta N_o S^{-1} \sum_{s=1}^{S} \lvert f_{g,s}^u \rvert^{-2}$ | $\beta N_o S^{-1} \sum_{s=1}^{S} \lvert f_{g,s}^u \rvert^{-2}$ |

Effectiveness of above algorithm depends on with changing CSI how the matrix $\{P_g^u\}$ is formed. Matrix $\{P_g^u\}$ will be the (G x U) matrix of which elements are dependent on CSI.

## 4. Simulation results

For carrying out simulations the environment is selected as given below in table 2. It is assumed that the CSI is available at the base station. The parameter describing CSI is the channel fading parameter ( f ), which is to be applied as input in the proposed algorithm and output is the no. of channels allocated for the given total transmit power ($P_T$ ).

TABLE II.
SIMULATION PARAMETERS

| Parameter | Environment |
|---|---|
| Total no. of channels (N) | 128 |
| Total no. of groups (G) | 8 |
| Total no.of users (U) | 8 |
| BER | $10^{(-2)}$ |
| Channel fading (f) | 0-12dB |
| Noise power spectral density (No) | 0.16 watts/Hz |
| Target SINR value (β) | -2ln5BER |

Frequency domain combining weights will be different for different combining schemes and accordingly throughput will vary for Minimal ratio combining (MRC), Equal gain combining (EGC) and Zero force combining (ZFC) schemes[1].

Simulations are carried out for the environment selected as in table 2. First combining scheme under consideration is MRC. For one set of input parameters(f) , the total power ($P_T$ ) is varied and the throughput obtained is shown in fig.4. Similarly for different sets of inputs we will get number of variations, for one of the set of such inputs a much better result is obtained as shown in fig.5. In both the cases as transmit power increases throughput increases. one can clearly notice that proposed algorithm gives significant improvement in the throughput obtained.

In fig. 6 throughput versus the maximum transmit power is obtained for EGC and ZFC schemes. As both combining schemes requires same amount of transmit power for one same channel, the throughput obtained is same for EGC and ZFC scheme as can be seen in fig.6. Comparatively in this case for EGC and ZFC combining schemes all the channels get allocated for a much less power for near about -9dB.

Fig.7 is intended for modified algorithm only for three combining schemes. One can observe that in this case with EGC, ZFC combining schemes much higher throughput is obtained at a very little power as compared to MRC scheme.

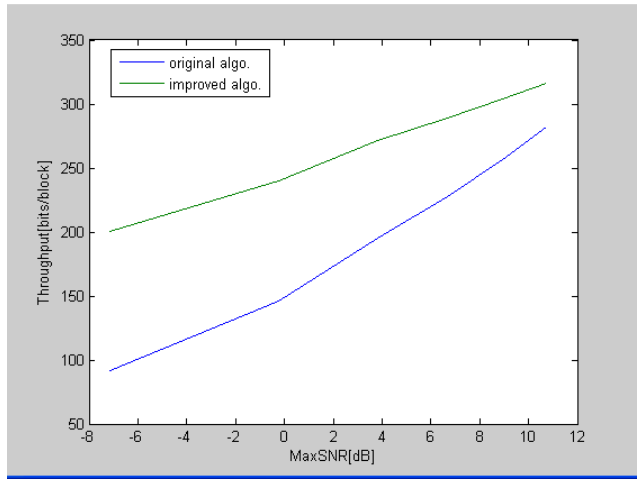

Figure.4. Throughput vs maximum transmit power for one set of input for MRC scheme.

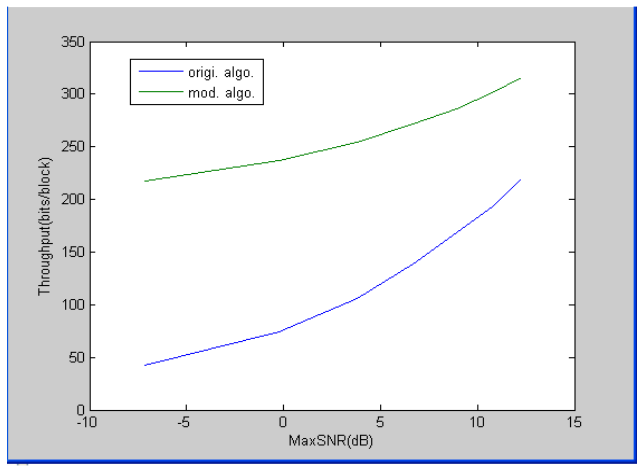

Figure.5.. Throughput vs maximum transmit power for second set of input for MRC scheme.

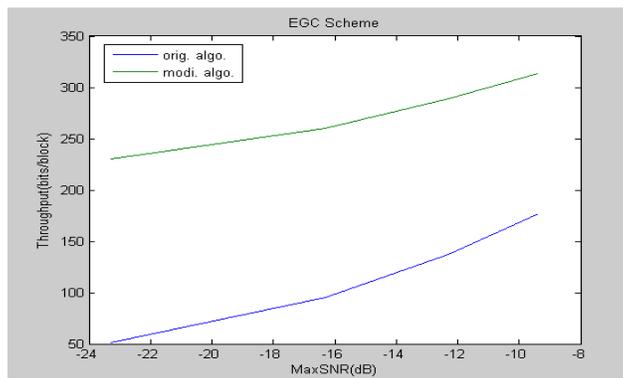

Figure.6.. Throughput vs maximum transmit power for EGC and ZFC scheme.

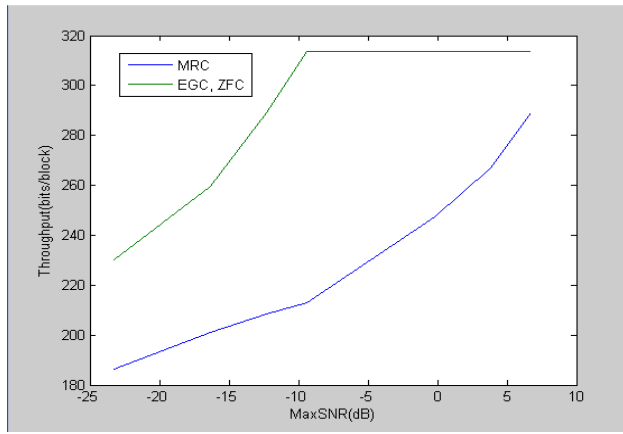

Figure.7. Throughput vs maximum transmit power power using modified algorithm.

## 5. CONCLUSIONS

In this paper an improved algorithm for throughput maximization is proposed. It gives an adaptive group assignment technique such that the available power at the base station will be utilized efficiently and maximum no. of channels will be allocated to the users.

Simulation results shows that the proposed algorithm gives much better results as compared to original one for all the three combining schemes. The future work can be carried out by first employing adaptive grouping of users and then applying proposed algorithm for group allocation.